# The Possibility of Investigating Ultra-High-Energy Cosmic-Ray Sources Using Data on the Extragalactic Diffuse Gamma-Ray Emission

A. V. Uryson*

*Lebedev Physical Institute, Russian Academy of Sciences, Leninskii pr. 53, Moscow, 119991 Russia*

*E-mail:uryson@sci.lebedev.ru

**Abstract**—We provide our estimates of the intensity of the gamma-ray emission with an energy near 0.1 TeV generated in intergalactic space in the interactions of cosmic rays with background emissions. We assume that the cosmic-ray sources are point like and that these are active galactic nuclei. The following possible types of sources are considered: remote and powerful ones, at redshifts up to $z = 1.1$, with a monoenergetic particle spectrum, $E = 10^{21}$ eV; the same objects, but with a power-law particle spectrum; and nearby sources at redshifts $0 < z \leq 0.0092$, i.e., at distances no larger than 50 Mpc also with a power-law particle spectrum. The contribution of cosmic rays to the extragalactic diffuse gamma-ray background at an energy of 0.1 TeV has been found to depend on the type of sources or, more specifically, the contribution ranges from $f << 10^{-4}$ to $f \approx 0.1$, depending on the source model. We conclude that the data on the extragalactic background gamma-ray emission can be used to determine the characteristics of extragalactic cosmic-ray sources, i.e., their distances and the pattern of the particle energy spectrum.

## INTRODUCTION

The sources of ultra-high-energy cosmic rays (UHECRs) have not been established at present. Information about the origin of UHECRs is usually obtained from their energy spectrum. In intergalactic space CRs lose their energy while interacting with the cosmic microwave background. As a result of the energy losses, CR particles with energies $E > 10^{20}$ eV will not reach the Earth if they come from distances greater than ~100 Mpc. Therefore, the measured CR energy spectrum can be cut off, i.e., there will be no particles with energies $E > 10^{20}$ eV (the GZK effect, see Greisen 1966; Zatsepin and Kuzmin 1966). If the particles arrive from smaller distances, then there is no cutoff in the energy spectrum. The spectra obtained from measurements at giant arrays, the Pierre Auger Observatory and the Telescope Array, are cut off. However, the spectrum cutoff can arise for a different reason: the particles are accelerated only to an energy of $10^{21}$ eV as a consequence of the conditions in their sources (Medvedev 2003; Uryson 2004). Since the UHECR sources are unknown, this possibility cannot be ruled out, and the reason for the spectrum cutoff remains unclear.

Apart from the GZK effect, there is another manifestation of the CR interactions with the cosmic microwave background: particles initiate electromagnetic cascades in intergalactic space (Hayakawa 1966; Prilutsky and Rozental 1970). The spectrum of the gamma-ray emission generated in cascades was studied theoretically by Prilutsky and Rozental (1970). Subsequently, the spectra of the cascade gamma-ray emission were calculated by Gavish and Eichler (2016) for several possible classes of UHECR sources uniformly distributed in redshift $z$ with a power-law particle spectrum in the sources and with a spectral index in the range 2.1–2.5; the spectra of the cascade gamma-ray emission calculated with such CR spectral indices were found to be virtually identical.

Gamma-ray emission is also generated in other processes. It is produced in CR interactions with interstellar matter and radiation in our Galaxy. In addition, gamma-ray photons are emitted by extragalactic discrete sources, for example, active galactic nuclei (AGNs) and gamma-ray bursts. Many of the discrete sources are so far unresolved. Finally, gamma-ray emission is generated in dark matter decays. Can the gamma-ray emission generated in cascades be selected from photons of a different nature? The extragalactic diffuse emission is separated from the gamma-ray emission. For this purpose, information about the Galactic gamma-ray emission is obtained from detailed models of the distribution of CR sources and matter in the Galaxy and the Galactic geometry; the contribution of unresolved sources can be estimated theoretically from the distribution of observed sources and their fluxes (see Ackermann et al. (2015) and references therein). Thus, the extragalactic gamma-ray emission was identified and measured by the OSO-3 (Clark et al. 1968), SAS-2 (Fichtel et al. 1975) satellites, and, subsequently, the Fermi Large Area Telescope (Fermi LAT) (Ackermann et al. 2015). The contribution of dark matter decays to the extragalactic diffuse gamma-ray emission is estimated theoretically (see, e.g., Gavish and Eichler (2016) and references therein). Given the intensity of the extragalactic diffuse gamma-ray emission and the contributions of the listed sources and processes to it, the intensity of the cascade gamma-ray photons can then be determined in principle.

Here, we estimated the intensity of the gamma-ray photons with an energy of 0.1 TeV generated in cascades for two classes of possible UHECR sources: Seyfert nuclei within ∼50 Mpc and BL Lac objects. We chose the energy of 0.1 TeV, because the extragalactic background light attenuates insignificantly the gamma-ray emission with such an energy (Dwek and Krennrich 2013): photons with an energy $E \approx 0.1$ TeV come from distances $z \approx 0.1$–$0.2$ with a flux attenuation of less than 10%; the gamma-ray flux that traversed the distances $z = 0.5$, $0.6$, and $1$ is attenuated by 20, 30, and 50–60%, respectively.

We found that the intensity of the gamma-ray photons with an energy of 0.1 TeV generated in cascades depends on the characteristics of their sources, i.e., their distances and the initial CR energy spectrum, and could account for ~10% of the extragalactic gamma-ray emission measured with Fermi LAT.

## INTERGALACTIC CASCADES

In intergalactic space UHECR protons interact with the microwave background: $p + \gamma_{rel} \rightarrow p + \pi^0$ and $p + \gamma_{rel} \rightarrow n + \pi^+$, where $\gamma_{rel}$ is a relic photon. The cross sections for these reactions $\sigma_{p\gamma}$ as a function of proton energy were given by the Particle Data Group (2004). The mean free paths of protons as a function of their energy and the energy transfer from the proton to the relic photon were calculated by Stecker (1968).

The development of a cascade is described in Hayakawa (1966) and Prilutsky and Rozental (1970): the pions produced in the interactions of CR protons with the relic radiation decay: $\pi^0 \rightarrow 2\gamma$, $\pi^+ \rightarrow \mu^+ + \nu$, and $\mu^+ \rightarrow e^+ + \nu + \nu$, while the generated gamma-ray photons and electrons interact with the background emission, relic and radio photons (Berezinsky 1970) in the reactions $\gamma + \gamma_b \rightarrow e^+ + e^-$ (pair production) and $e + \gamma_b \rightarrow e' + \gamma'$ (inverse Compton scattering). Here, the index b and the prime denote the background photon and the scattered particle, respectively. The cross sections for these reactions are given in the book by Ginzburg (1979).

The pair production is a threshold process; it is possible if the photon energy $E_\gamma > E_t$, where $E_t$ is the threshold energy:

$$E_t = (mc^2)^2 / \varepsilon_b, \qquad (1)$$

where $mc^2$ is the electron mass, $mc^2 = 0.5$ MeV, and $\varepsilon_b$ is the background photon energy. The pair production cross section $\sigma_{\gamma\gamma}$ is

$$\sigma_{\gamma\gamma} \approx 3/8 \sigma_T a^2 / \{[2+2a^2 - a^4] \ln(a^{-1} + (a^{-2} - 1)^{1/2}) - (1 - a^2)^{1/2}(1 + a^2)\}, \qquad (2)$$

where $a = mc^2 / E_{ec}$, $E_{ec}$ is the center-of-mass energy of the photons:

$$E_{ec} = [E_e \varepsilon_b (1 - \cos \psi)]^{1/2}, \qquad (3)$$

$\psi$ is the angle between the photon momenta in the observer's frame.
The inverse Compton scattering cross section $\sigma_{IC}$ $\sigma_{IC}$ at $E_e > E_t$ is

$$\sigma_{IC} \approx 3/8 \sigma_T (mc^2)^2 / (E_e \varepsilon_b) \{\ln[2 E_e \varepsilon_b / (mc^2)^2] + 0.5\}. \qquad (4)$$

At $E_e < E_t$ the cross section (4) is equal to the Thomson one, $\sigma_{IC} = \sigma_T \approx 6.65 \times 10^{-25}$ cm$^2$, and the electron scatters comparatively soft photons with a mean energy

$$E_\gamma = 4/3 \varepsilon_b (E_e / mc^2)^2. \qquad (5)$$

The higher-order reactions with background photons $\gamma + \gamma_b \rightarrow e^+ + e^- + e^+ + e^-$ (the production of two electron–positron pairs) and $e + \gamma_b \rightarrow e' + e^+ + e^-$ (triplet pair production in an electron–photon interaction) are suppressed, make a very small contribution to the cascade process (Bhattacharjee and Sigl 2000), and we disregarded them.

Apart from the inverse Compton scattering, electrons produce photons via synchrotron radiation in magnetic fields. The intensity of synchrotron radiation depends on the magnetic field strength. It can be intense enough to lead to substantial electron energy losses and to stop the cascade process. In intergalactic space the magnetic fields are too weak to disrupt the development of a cascade. We discussed this previously (Uryson 2006).

In the Galaxy the magnetic field is, on average, $B \approx 3 \times 10^{-6}$ G. In such a field the energy loss by UHE electrons through synchrotron radiation is faster than the inverse Compton scattering. Therefore, we assume that the cascade ceases to develop in the Galaxy.

CRs fly in an expanding Universe and, hence, lose their energy adiabatically. Traversing the distance from a point with redshift $z_i$ to a point with redshift $z_{i+1}$, the particle loses an energy $(\Delta E)_{ad}=(Ez_{i+1}-Ez_i)/(1+z_i)$.

DESCRIPTION OF THE MODEL

Our model assumptions concern three points: the CR sources, the background emission in intergalactic space, and the extragalactic magnetic fields.

We assume the CR sources to be pointlike. This assumption is consistent with results from Abu-Zayyad et al. (2013). As UHECR sources we consider AGNs: BL Lac objects at redshifts up to $z = 1.1$, and moderate-luminosity Seyfert nuclei at redshifts $0 \leq z \leq 0.0092$, i.e., within $\sim 50$ Mpc. These objects were analyzed as possible UHECR sources by Kardashev (1995), Uryson (1996, 2001), and Gorbunov et al. (2002).

We obtained the spatial distributions of these sources previously (Uryson 2006) based on the redshifts of AGNs from the catalogue by Veron-Cetty and Veron (2003). The peak in the $z$ distribution is in the range $z \approx 0.2$–$0.3$ and at $z \approx 0.003$ for the BL Lac objects and Seyfert nuclei under consideration, respectively.

We assume that the distance to a source $r$ is related to its redshift $z$ as

$$r=2/3cH^{-1}(1-(1+z)^{-3/2}) \text{ [Mpc]}, \tag{6}$$

at $z > 0.1$ (in accordance with the Einstein–de Sitter model with $\Omega = 1$) and as

$$r = cz/H \text{ [Mpc]} \tag{7}$$

at lower $z$, where $c$ is the speed of light and $H$ is the Hubble constant.

We consider the cases where CRs in sources are accelerated either by electric fields or at shock fronts. The initial CR spectrum is monoenergetic and exponential in the former and latter cases, respectively. We assumed the initial spectrum in BL Lac objects to be monoenergetic with $E = 10^{21}$ eV or exponential with an index $\chi$ in the range $\chi = 2$–$3$; the initial spectrum in Seyfert nuclei is exponential with an index $\chi = 2\text{–}3$ (Uryson 2004). CRs in sources are accelerated to a maximum energy of $10^{21}$ eV (Medvedev 2003; Uryson 2004).

In what follows, the BL Lac objects and Seyfert nuclei will be called remote and nearby sources, respectively.

Thus, the model assumptions about the UHECR sources are the distances to them and the initial spectra of accelerated particles. In total we consider three types of sources: remote sources with an initial monoenergetic CR spectrum, remote sources with an exponential CR spectrum, and nearby sources with an exponential CR spectrum.

We assume that the UHECRs consist of protons.

The model assumptions about the background emission were the following. The microwave background has a Planck energy distribution with a mean energy $\varepsilon_{rel} = 6.7 \times 10^{-4}$ eV, the mean photon density is $n_{rel} = 400$ cm$^{-3}$. The high-energy "tail" in the distribution is formed by photons with energy $\varepsilon_t = 1 \times 10^{-3}$ eV, their mean density is $n_t = 42$ cm$^{-3}$. The anisotropy in the microwave background does not affect the development of intergalactic cascades and was disregarded.

The spectrum of the extragalactic background radiation was measured by Clark et al. (1970) and was then studied theoretically by Protheroe and Biermann (1997). In the model we adopted the results from Protheroe and Biermann (1997) by taking into account the evolution of radio sources: the radio background exists at energies up to $\varepsilon_b \approx 4 \times 10^{-10}$ eV, the density of radio photons at this energy is $n_b \approx 1$ cm$^{-3}$.

We neglect the electron energy losses in intergalactic magnetic fields.

## CALCULATIONS

The goal of our calculations is to determine the number of photons with energy $E_\gamma \approx 0.1$ TeV that were produced in the cascade process initiated by a proton coming from the source to the Galaxy.

First we calculated the total number of gamma-ray photons $N_\gamma$ with energy $E_\gamma \approx 100$ TeV produced by one proton coming from the source to the Galaxy. The values of $N_\gamma$ for different models were obtained as follows.

The distances to the UHECR sources were calculated in accordance with their spatial distribution: we generated the redshift $z$ of a source by the Monte-Carlo technique and then calculated its distance. Subsequently, having specified the mean free path of a proton in its interactions with the background radiation, $<L> = 1/(n_{rel}\sigma_{p\gamma})$, we calculated the proton mean free path $L$ by the Monte Carlo technique. The proton energy losses $\Delta E_{ad}$ due to the expansion of the Universe were calculated at the point of proton–background photon interaction. The proton energy after its interaction with the photon was then calculated. We took into account the energy dependences of the interaction cross section and the inelasticity coefficient from Stecker (1968). The procedure was repeated until the proton reached the Galaxy or its energy decreased to $E < 4\times10^{19}$ eV (because the mean free paths for protons of such energies are large, several hundred Mpc).

Gamma-ray photons with energy $E_\gamma=0.5E_{\pi0}$, where $E_{\pi0}$ is the $\pi^0$-meson energy, are produced in neutral pion decays.

The pair production processes were considered as follows. First we generated the angle between the momenta of the cascade and background photons in the observer's frame of reference by the Monte-Carlo technique and then calculated the center-of-mass energy remained in the cascade photon. Subsequently, we calculated the pair production cross section $\sigma_{\gamma\gamma}$ and the mean free path of the cascade photon in this process $\lambda_{\gamma\gamma}=1/(n_b\sigma_{\gamma\gamma})$ for each type of background photon, the relic and radio ones. We chose the interaction with the photon for which the mean free path was shorter. We assumed that one of the particles in the electron–positron pair had energy $E_t$, while the energy of the other one was ($E_\gamma-E_t$).

The inverse Compton scattering was considered similarly to the pair production. The cascade electron involved in the scattering was traced until its energy decreased to 100 TeV (because at lower energies the electron loses its energy comparatively slowly while scattering soft photons).

This scheme was used to calculate the total number of gamma-ray photons $N_\gamma$ with energy $E_\gamma \approx$ 100 TeV produced by one proton coming from the source to the Galaxy.

The procedure was repeated for 1000 sources, each of which emitted one proton. The mean number of cascade gamma-ray photons with energy $E_\gamma \approx$ 100 TeV was then calculated.

Gamma-ray photons with such an energy do not reach the Earth due to their interactions with the extragalactic background light. We assumed that photons with $E_\gamma \approx$ 100 TeV interacted with the background light until their energy decreased to $E_{\gamma 1} \approx 0.1$ TeV. Photons with such an energy propagate in intergalactic space with insignificant absorption (Dwek and Krennrich 2013). Next, we assumed that in its interactions with the background light one gamma-ray photon produced $M_\gamma = 10^3$ photons with energy $E_{\gamma 1}= 0.1$ TeV, where $M_\gamma= E_\gamma/E_{\gamma 1}$. As a result, while propagating from the source to the Galaxy, a CR proton generates $N_\gamma M_\gamma$ photons at energy $E_{\gamma 1} = 0.1$ TeV, and these photons reach the array.

RESULTS

In the model of remote sources with a monoenergetic initial CR spectrum, protons produce $N_\gamma \approx 10^6$ gamma-ray photons with energy $E_\gamma \approx 100$ TeV on their way to the Galaxy. As a result, $M_\gamma = 10^9$ photons with energy $E_{\gamma 1} = 0.1$ TeV are produced.

In the models of sources with an exponential initial CR spectrum the number of cascade photons is much smaller. The protons emitted by remote sources produce $N_\gamma \approx 1.5 \times 10^3$ gamma-ray photons with energy $E_\gamma \approx 100$ TeV on their way to the Galaxy, which generate $M_\gamma = 1.5 \times 10^6$ photons with $E_{\gamma 1} = 0.1$ TeV. For the protons from nearby sources (Seyfert nuclei) the number of produced gamma-ray photons is $N_\gamma \approx 0$, $M_\gamma \approx 0$. $N_\gamma$ and $M_\gamma$ do not depend on the spectral index of the initial exponential CR spectrum that we considered, $\chi = 2$ or $3$.

Let us estimate the intensity $I\gamma$ of the gamma-ray emission with an energy of 0.1 TeV generated in cascades. Since the CR spectra fall steeply, for our estimates we will use approximations of the integrated intensity:

$$I_\gamma (\geq 0.1 \text{ TeV}) = M_\gamma \, I_{CR}(E > 4 \times 10^{19} \text{ eV}), \qquad (8)$$

where $I_{CR}$ is the UHECR intensity. An approximation of the integrated CR intensity is given in the book by Berezinsky et al. (1990):

$$I_{CR}(>E) = 1 \times (E \text{ (GeV)})^{-1.7} \text{ particles/(cm}^2 \text{ s sr)} \qquad (9)$$

at $10$ GeV $< E < 3 \times 10^6$ GeV;

$$I_{CR}(>E) = 3 \times 10^{-10} (E \text{ (GeV)}/10^6)^{-2.1} \text{ particles/(cm}^2 \text{ s sr)} \qquad (10)$$

at $E > 3 \times 10^6$ GeV.
Hence we obtain

$$I_{CR}(E > 4 \times 10^{19} \text{ eV}) \approx 6 \times 10^{-20} \text{ (cm}^2 \text{ s sr)}^{-1}. \qquad (11)$$

Using the estimate (13), we find from (8) that the intensity of the cascade gamma-ray emission in the adopted models is:

—in the model of remote sources with a monoenergetic CR spectrum

$$I_\gamma (E_\gamma \geq 0.1 \text{ TeV}) \approx 6 \times 10^{-11} \text{ (cm}^2 \text{ s sr)}^{-1}, \qquad (12)$$

—in the model of remote sources with an exponential CR spectrum

$$I_\gamma (E_\gamma \geq 0.1 \text{ TeV}) \approx 6 \times 10^{-14} \text{ (cm}^2 \text{ s sr)}^{-1}, \qquad (13)$$

—in the model of nearby sources

$$I_\gamma (E_\gamma \geq 0.1 \text{ TeV}) \ll 6 \times 10^{-14} \text{ (cm}^2 \text{ s sr)}^{-1}. \qquad (14)$$

The intensity of the diffuse extragalactic gamma-ray emission from Fermi LAT data (Ackermann et al. 2015) is

$$I_{EGB} (>0.1 \text{ TeV}) \approx 7 \times 10^{-10} \text{ (cm}^2 \text{ s sr)}^{-1}. \qquad (15)$$

From (12) – (15) we estimate the relative contribution $f$ of cascade gamma-ray photons to the extragalactic gamma-ray background at energy $E_{\gamma 1} \approx 0.1$: $f \ll 10^{-4}$ if the UHECR sources are nearby ones; $f \approx 10^{-4}$ if the UHECR sources are remote ones, also with an exponential particle energy spectrum; and $f \approx 0.1$ if the UHECR sources are remote ones, but with a monoenergetic CR spectrum.

DISCUSSION

There are two reasons for the difference in the number of cascade gamma-ray photons $N_\gamma$: the resonance in the energy dependence of the proton–relic photon (p-$\gamma_{rel}$) interaction cross section and the shape of the initial CR spectrum. The emitted proton interacts with the microwave background on its way from the source until its energy decreases to $4 \times 10^{19}$ eV. The mean free path of the proton is then hundreds of Mpc, and the probability of its interaction with the background radiation becomes low. As a consequence, the proton emitted by a source with a monoenergetic spectrum at an energy of $10^{21}$ eV interacts with the background radiation $\sim 10$ times on its way to the Galaxy, initiating ten electromagnetic cascades. In sources with exponential spectra the overwhelming majority of protons are emitted with an energy of $5 \times 10^{19}$ eV, irrespective of the spectral index. Protons with such an energy interact with the background radiation one or two times on their way to the Galaxy, giving rise to one or two cascades. In addition, the distances to nearby sources are too short for a cascade to develop. The energy dependence of the proton mean free path is shown in Fig. 1.

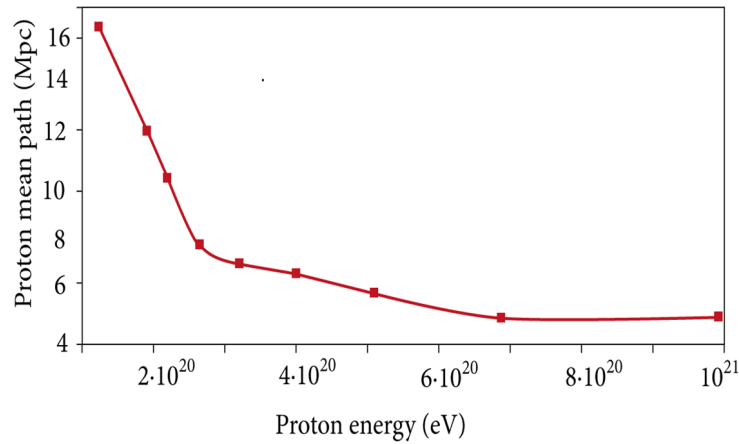

**Fig. 1.** Energy dependence of the proton mean free path in its interactions with the background radiation in intergalactic space.

The energy dependence of the photon mean free path during the pair production is shown in Fig. 2; the energy dependence of the electron mean free path upon the inverse Compton scattering is shown in Fig. 3. At the energies under consideration the cascade electron and photon mean free paths do not exceed 10 Mpc. Therefore, the interactions of cascade electrons and photons with the background radiation do not affect the difference in $N_\gamma$.

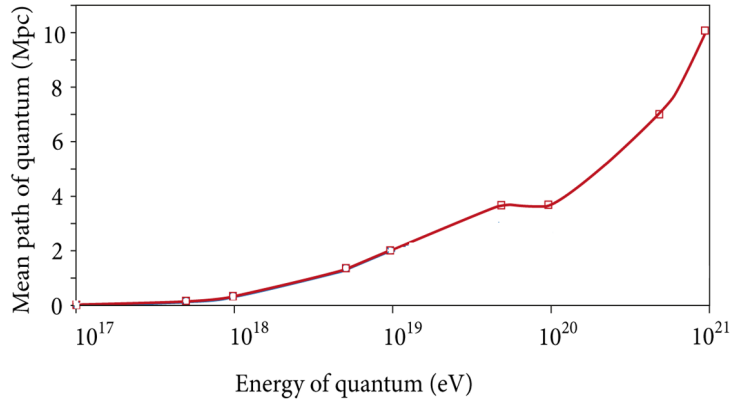

**Fig. 2.** Energy dependence of the photon mean free path during the pair production.

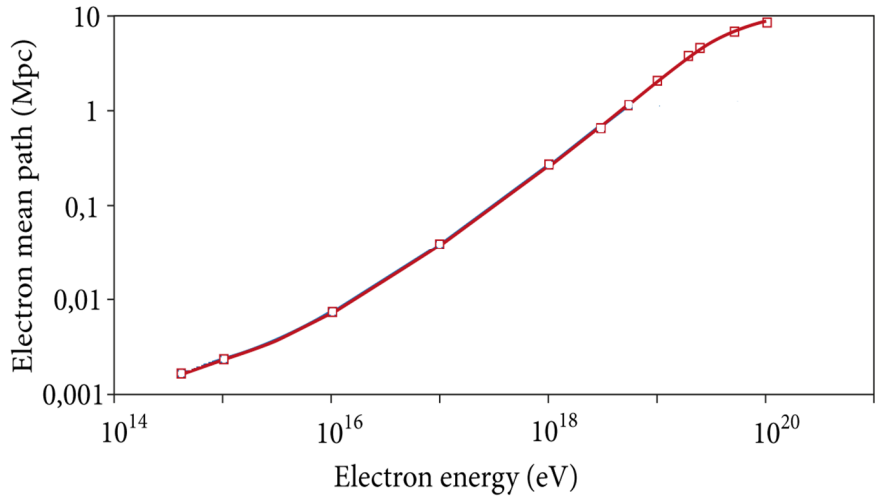

**Fig. 3.** Energy dependence of the electron mean free path during the inverse Compton scattering.

CONCLUSIONS

We analyzed the gamma-ray emission produced in electromagnetic cascades when UHECRs propagate in intergalactic space. The following types of possible CR sources were considered: powerful remote sources at redshifts up to $z = 1.1$, i.e., BL Lac objects hundreds of Mpc away, with an initial monoenergetic particle spectrum, $E = 10^{21}$ eV; remote sources, also BL Lac objects, but with a power-law initial particle spectrum; and nearby sources at $0 < z \leq 0.0092$, i.e., within ~50 MPc, Seyfert nuclei, with a power-law initial particle spectrum. We chose the types of sources based on the results from Uryson (2006). For these sources we estimated the intensity of the gamma-ray emission at an energy of 0.1 TeV generated in cascades. We chose such an energy, because gamma-ray photons with this energy are attenuated insignificantly by the extragalactic background light (Dwek and Krennrich 2013). The relative contribution $f$ of the cascade gamma-ray emission with an energy of 0.1 TeV to the extragalactic diffuse gamma-ray background turned out to depend on the characteristics of CR sources. The contribution of cascade gamma-ray photons is negligible, $f \approx 10^{-4}$ and $f \ll 10^{-4}$ if the CR spectrum in the sources is exponential (for the BL Lac objects and nearby Seyfert nuclei, respectively); the contribution is $f \approx 0.1$ if the UHECR sources are BL Lac objects with a monoenergetic initial particle spectrum. We obtained these estimates using the Fermi LAT data (Ackermann et al. 2015). The distributions of sources in redshift $z$ were derived from the data of the catalogue by Veron-Cetty and Veron (2003).

Therefore, by separating the contribution of cascade gamma-ray photons from the diffuse extragalactic gamma-ray emission, we can determine the type of UHECR sources and the pattern of the particle spectrum in the sources: whether these are BL Lac objects with a monoenergetic CR spectrum or not. Only in this case is the contribution of cascade gamma-ray photons appreciable: $\sim 10\%$.

Detailed calculations of the intensity and spectrum of the cascade gamma-ray emission will be performed using the TransportCR code (Kalashev 2003). Refined data on the extragalactic diffuse gamma-ray background will be obtained in the GAMMA-400 experiment (Topchiev et al. 2016).

## ACKNOWLEDGMENTS

I thank N.P. Topchiev, who read the manuscript of the paper and made remarks on it. I thank the referee for the discussion and critical remarks.

*Translated by V. Astakhov*